\documentclass{article}
\usepackage[utf8]{inputenc}

\usepackage{alphabeta}
\usepackage{amssymb}
\usepackage[T1]{fontenc}    
\usepackage{hyperref}       
\usepackage{url}            
\usepackage{booktabs}       
\usepackage{amsfonts}       
\usepackage{nicefrac}       
\usepackage{microtype}      
\usepackage{amsmath}
\usepackage{amssymb}
\usepackage{amsmath, amsthm}
\usepackage{graphicx}
\usepackage{color}
\usepackage[colorinlistoftodos]{todonotes}
\usepackage{datetime}
\theoremstyle{definition}

\title{Unveiling the Evolution of Mobile Networks: From 1G to 7G}

\author{Ellie Zontou\\
  Dept. of Computer Engineering and Informatics\\
  University of Patras\\
  Patras, 26504, Greece \\
  \texttt{ezontou@ceid.upatras.gr} \\
}

\begin{document}
\maketitle

\begin{abstract}
The evolution of cellular networks has played a pivotal role in shaping the modern telecommunications landscape. This paper explores the journey of cellular network generations, beginning with the introduction of Japan's first commercial 1G network by Nippon Telegraph and Telephone (NTT) Corporation in 1979. This analog wireless network quickly expanded to become the country's first national 1G network within a remarkably short period.

The transition from analog to digital networks marked a significant turning point in the wireless industry, enabled by advancements in MOSFET (Metal-Oxide-Semiconductor Field Effect Transistor) technology. MOSFET, originally developed at Bell Labs in 1959, underwent modifications to suit cellular networks in the early 1990s, facilitating the shift to digital wireless mobile networks. The advent of the 2G generation brought forth the first commercial digital cellular network in 1991, sparking recognition among manufacturers and mobile network operators of the importance of robust networks and efficient architecture. As the wireless industry continued to experience exponential growth, the significance of effective network infrastructure became increasingly evident.

In this research, our aim is to provide a comprehensive overview of the entire spectrum of cellular network generations, ranging from 1G to the potential future of 7G. By tracing the evolution of these networks, we aim to shed light on the transformative developments that have shaped the telecommunications landscape and explore the possibilities that lie ahead in the realm of cellular technology.
\end{abstract}

\small{\it }

\section{Introduction}

The cellular technology forms the foundation of mobile phone networks, earning them the common name "cell phones." Instead of relying on a single large transmitter, cellular technology utilizes multiple small transmitters linked together. These cellular networks serve as high-speed voice and data communication networks, capable of accommodating cellular devices with advanced multimedia features and seamless roaming capabilities \cite{Peng}. Over time, cellular networks have become the cornerstone of the communications industry.

\begin{figure}[ht!]
\centering\includegraphics[width=7cm]{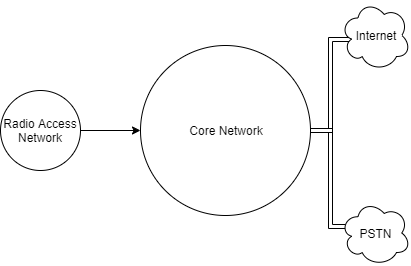}
\caption{A cellular network's architecture}
\label{fig:Figure 1}
\end{figure}

As illustrated in \cite{Peng} and depicted in Figure~\ref{fig:Figure 1}, subscribers gain access to a cellular network through radio signals facilitated by a radio access network. This component of the mobile telecommunication system employs radio access techniques to connect specific devices to other network elements, particularly the Core Network. The Core Network undertakes essential operations such as traffic routing and handling subscriber requests. Additionally, it serves as the link between the cellular network and other networks like the Public Switched Telephone Network (PSTN) and the Internet. The PSTN allows users to make landline calls, encompassing various telephone networks worldwide. Furthermore, Internet connectivity empowers the cellular network to provide cutting-edge multimedia services.

In summary, a cellular network consists of discrete areas known as cells or cell sites, strategically dispersed across a wide geographic area. Each cell site is equipped with one or more transceivers that provide radio coverage for the respective region, typically housed within a base station. To prevent interference and ensure satisfactory service quality, each cell site operates on different frequencies than nearby cells. By positioning cells in close proximity to one another, the network can offer radio coverage over a large area and facilitate seamless communication for mobile devices transitioning between cells, such as a cell phone in a moving vehicle. The range of a cell can vary significantly depending on factors such as configuration, size, and environmental conditions, spanning from approximately 10 meters to over 25 miles. In larger areas, cell repeaters can be deployed to enhance coverage.

As each new generation of cellular networks emerges, bringing with it evolving requirements, technologies, and solutions, significant advancements have occurred. Cellular networks have evolved into the fundamental infrastructure for various applications, including the Internet of Things (IoT) devices and robotics. Presently, with the emergence of 6G and 7G on the horizon, new possibilities and opportunities are beginning to unfold.

\section{The Cellular network generations}
\subsection{First Generation (1G)}
The advent of analog telecommunications standards marked the era of 1G, which primarily focused on delivering foundational voice services. Japan played a pioneering role in the late 1970s by establishing the first mobile network in Tokyo. This was followed by the deployment of NMTs (Nordic Mobile Telephones) in Europe and the introduction of AMPS (Advanced Mobile Phone Service) technology. However, these technologies faced certain limitations. AMPS, for instance, had inherent restrictions in terms of available channels and the associated devices were relatively expensive. Moreover, users encountered challenges such as suboptimal voice quality, capacity constraints, limited security measures, and slow data speeds.

One of the major issues with 1G technology was its reliance on analog impulses for information transmission. Analog signals were less effective compared to digital signals, leading to reduced efficiency in communication. This limitation became increasingly apparent as technology advanced and the demand for more sophisticated services grew.

Despite these challenges, the introduction of 1G networks laid the foundation for the remarkable advancements that followed. It set the stage for the subsequent generations of mobile networks, which addressed the shortcomings of their predecessors. The transition from analog to digital technology brought about significant improvements in voice quality, data capacity, security features, and transmission speeds.
\subsection{Second Generation (2G)}

The GSM (Global System for Mobile Communication) technology forms the foundation for 2G mobile networks. One of the key features of GSM is terminal mobility, which allows mobile users to move between different locations while maintaining continuous connectivity. This mobility is facilitated by a subscriber identity module (SIM) inserted into the GSM network, which stores the unique number assigned to the mobile user \cite{Rahnema}. The SIM card plays a crucial role in enabling personal mobility within the GSM network.

The GSM network comprises four main components that work together to provide seamless operation: the mobile device itself, the Base Station Subsystem (BSS), the Network Switching Subsystem (NSS), and the Operation and Support Subsystem (OSS). The BSS consists of the Base Transceiver Station (BTS) and the Base Station Controller (BSC). The NSS includes essential elements such as the Mobile Switching Centre (MSC), Visitor Location Register (VLR), Home Location Register (HLR), Authentication Centre (AC), and Equipment Identity Register (EIR). These components collaborate to ensure the smooth functioning of the GSM network.

\begin{figure}[ht!]
\centering\includegraphics[width=7cm]{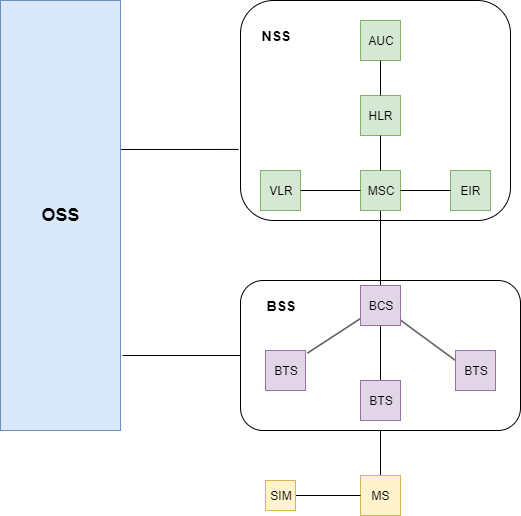}
\caption{GSM architecture}
\label{fig:Figure 2}
\end{figure}

Unlike the analog radio signals used in 1G networks, 2G networks introduced the use of digital radio signals. This digitalization allowed for multiple users to be accommodated on a single channel through a process called multiplexing. As a result, 2G networks enabled cell phones to be used for both voice and data communication.
The second generation (2G) of wireless mobile communication systems, driven by groundbreaking technology and the services it offered, was a resounding success. Figure ~\ref{fig:Figure 2} provides an overview of the GSM architecture, illustrating how 2G networks operate. 

\subsection{Third Generation (3G)}
The third generation of cellular networks brought significant advancements in data speed and network capabilities, enabling a wide range of services such as video calling, video streaming, gaming, and fast internet browsing. The Universal Mobile Telecommunications System (UMTS) was established as a result, combining and enhancing the capabilities of second-generation Global System Mobile (GSM) networks to enable international roaming \cite{Mishra, Holma}. UMTS, being a part of the 3G family, offers higher data speeds compared to its predecessors.

Under the International Telecommunication Union (ITU) IMT-2000 standard, 3G systems are classified as capable of supporting high-speed data ranging from 144 kbps to over 2 Mbps \cite{Mishra}. IMT-2000 serves as a global standard for 3G systems. Furthermore, 3G systems employ five key radio technologies, including Time Division Multiple Access (TDMA), Code Division Multiple Access (CDMA), and Frequency Division Multiple Access (FDMA), which facilitate efficient communication \cite{Holma}.

TDMA systems divide the channel time into frames, with each frame further divided into time slots, allowing only one user to send or receive data in each slot. CDMA, on the other hand, multiplies the narrowband message signal by a spreading signal (code) with a wide bandwidth before modulation and transmission, allowing for multiple users to share the same frequency band through spreading. FDMA assigns a specific frequency band or channel to each user, ensuring that no other user can utilize the same band during a call.

In conclusion, 3G systems have the capability to coexist with 2G technologies, enabling seamless transition and compatibility. These systems are designed to accommodate future growth and expansion with minimal costs, aiming to provide broader coverage and enhanced services \cite{Mogal}.

\subsection{Fourth Generation (4G)}

The advancement of next-generation systems, commonly referred to as 4G, became imperative as the existing IMT-2000 standard failed to address the challenges related to higher data rates and capacity. A 4G system offers users significantly higher data rates for voice, data, and streaming multimedia compared to 3G or 2G systems. To achieve faster user data rates and lower latency, 4G leverages research conducted in the fields of GSM, EGPRS, WCDMA, and HSPA. With transmission rates exceeding 20 Mbps, 4G technology delivers impressive data transfer capabilities. Additionally, 4G networks exhibit significantly lower latency, which is crucial for real-time interactions like video conferencing, where minimal delay is essential. The transmission and reception capabilities of 4G are enabled by two key technologies: Orthogonal Frequency Division Multiplexing (OFDM) and Multiple Input Multiple Output (MIMO).

OFDM serves as the foundation for 4G systems by dividing the available spectrum into several sub-channels. Each of these narrowband sub-channels experiences nearly flat fading, simplifying the equalization process. The frequency responses of these sub-channels are orthogonal and overlapping, providing high spectral efficiency and justifying the name "Orthogonal Frequency Division Multiplexing" \cite{Edfors}. OFDM offers faster speeds compared to the primary 3G technologies. On the other hand, MIMO is a 4G LTE antenna technology that utilizes multiple antenna components at both the transmitter and receiver ends to enhance data rates and improve signal quality. The implementation of MIMO technology in 4G systems contributes to the overall performance improvement and efficiency in wireless communications.

These technological advancements in 4G networks have revolutionized the way users experience mobile communication, enabling high-speed data transfer, low-latency applications, and enhanced multimedia services.

\subsection{Fifth Generation (5G)}

The emergence of 5G cellular networks stems from the limitations of previous generations, such as 4G, in meeting the increasing demands for high-speed and reliable connectivity. While 4G networks initially brought about a significant leap in data rates, the International Telecommunication Union (ITU) recognized the need for a new standard to address the evolving requirements of mobile communication. This led to the definition of the fifth generation (5G) as "International Mobile Telecommunication 2020" (IMT-2020) \cite{Banda}.

The COVID-19 pandemic has further underscored the significance of 5G networks. With the widespread adoption of remote work, online education, and video conferencing, there has been a massive surge in mobile data traffic. The need for seamless video streaming, high-quality video conferencing, and real-time communication has become more critical than ever. 5G networks are designed to handle this increased demand by offering higher data rates, accommodating a larger number of wireless connections, reducing communication latency, and improving energy efficiency \cite{IMT}.

At the architectural level, the 5G core network introduces several key advancements. One notable aspect is the emphasis on control-plane/user-plane separation. By separating these two planes, the network can optimize resource allocation, improve scalability, and enhance flexibility in managing network services. This architectural approach enables efficient handling of control signaling and data traffic, leading to improved overall network performance.

\begin{figure}[h!]
\centering\includegraphics[width=8cm]{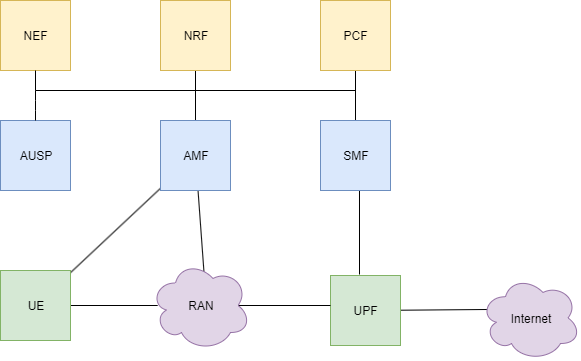}
\caption{5G's core network's architecture}
\label{fig:Figure 3}
\end{figure}

The 5G core network is designed based on a service-based architecture, which allows for the creation, deployment, and orchestration of network services in a more modular and flexible manner. This service-based approach enables efficient management of network functions, easier integration of new services, and enhanced customization according to specific application requirements. Figure ~\ref{fig:Figure 3} provides a high-level visualization of the 5G core network, illustrating the service-based architecture and the various network components involved \cite{Aranda}.

In summary, the transition to 5G networks is driven by the need for advanced connectivity capabilities that can meet the growing demands of modern applications and services. The COVID-19 pandemic has further accelerated the adoption of 5G due to the increased reliance on remote communication and digital connectivity. With its higher data rates, increased wireless capacity, reduced latency, and improved energy efficiency, 5G represents a significant leap forward in cellular network technology, enabling a wide range of innovative applications across industries and sectors.

The User-Plane Function (UPF) in 5G plays a crucial role in connecting the actual data flowing through the Radio Area Network (RAN) to the Internet. Its primary task is to handle the data traffic, ensuring its seamless transmission and reception. On the other hand, the control-plane functions comprise various components, with one of them being the Session Management Function (SMF). The SMF takes charge of managing user sessions, from their establishment to modification and termination, while also allocating IP addresses for IP PDU (Protocol Data Unit) sessions.

Another essential component is the Access and Mobility Management Function (AMF), which handles connection and mobility management activities. Although it gathers all connection- and session-related data from the User Equipment (UE), it is responsible for managing the connections and movements of UEs throughout the network. The core network also includes other functions such as the Network Exposure Function (NEF), Network Repository Function (NRF), and Policy Control Function (PCF), each serving specific roles in the overall network architecture.

As we look at the evolution of cellular networks from 1G to 5G, it becomes evident why it is essential to keep advancing. The increasing demands for faster data rates, higher capacities, low latency, and improved connectivity have been driving the continuous development of cellular technologies. This evolution has led researchers to explore and investigate the possibilities of the next generation, 6G. As we move towards the future, research and development efforts are already underway to envision and shape the cellular networks of 6G, which are expected to push the boundaries of wireless communication even further.

\subsection{Sixth Generation (6G)}

The objective of this chapter is to conduct a comprehensive examination of the characteristics of the upcoming generation of mobile networks, drawing on existing research documented in the bibliography.
6G wireless networks are poised to deliver even greater bandwidth compared to 5G, along with significantly lower latency. It is projected that 6G networks will gain widespread usage around 2030, and prominent telecommunications companies like Ericsson, Samsung, and Nokia are already spearheading research efforts in this area.

The exponential growth of big data in everyday life highlights the imperative to transition to a new generation of networks. The handling of today's data, characterized by its large volume and often real-time processing requirements, poses challenges. Moreover, this data is frequently semi-structured or unstructured. By 2030, it is expected that an advanced digital society will rely on ubiquitous wireless access. These factors serve as driving forces behind the initiation of 6G research \cite{Berndt}.

The 6G mobile network is anticipated to utilize air fiber technology, employing masts and transceivers installed on tall buildings to establish local networks that can provide remarkable speeds. This approach differs from relying solely on average performance figures. The combination of air fiber technology and local networks will enable the secure transmission of data from transmitters to destinations \cite{Ribeiro}. In its initial implementation, the 6G network will leverage the existing design and advantages of 5G, including wider frequency bands and an improved decentralized network architecture \cite{Alraih}. Therefore, it is reasonable to assume that millimeter waves, which are extremely high frequencies in the radio spectrum, will be employed, as in 5G. The availability of ample spectrum resources justifies the utilization of such frequencies in both 5G and 6G, with numerous unlicensed bands being available.

The fifth generation of cellular networks is characterized by its cloud-based nature. Looking ahead, during the design, deployment, and operation of 6G networks, machine learning (ML) and artificial intelligence (AI) are expected to play pivotal roles. Intelligent, self-organizing, and cost-effective 6G networks are necessary, and ML offers pragmatic solutions that can reshape the future of wireless network technologies. The integration of AI and ML aims to generate new revenue streams and enhance network performance. Deep learning AI approaches have already started replacing conventional algorithms, leading to significant reductions in power consumption and improved system performance \cite{Patil}.

To further illustrate the potential of AI in 6G, consider the example of self-driving cars. These autonomous vehicles have faced challenges in applying AI and ML algorithms to real-world situations, hindering their ability to learn effectively. In addition to addressing this problem, autonomous vehicles need to be aware of their location, surroundings, and other road users to compute routes and arrival times. This necessitates the rapid construction of on-the-fly networks while also being part of larger networks. Processing the immense volume of real-time data involved requires extremely low latency, even lower than what the 5G network can provide. It is evident that 6G will revolutionize daily life, particularly in applications such as self-driving cars, addressing significant challenges from their troubled past.

\subsection{Seventh Generation (7G) and onwards}

Looking ahead, the question arises: what lies beyond 6G? Despite 6G not yet being fully deployed, research on the subsequent generation has already commenced. Given the rapid transitions from 1G to 5G, it appears inevitable that a more advanced network will follow 6G. So, how might 6G be further improved?

The forthcoming intelligent cellular technology, known as 7G, is set to succeed 5G and 6G, promising significantly enhanced capacity, even higher frequencies, and vastly lower latency. Anticipated to meet the requirements of ultra-high bandwidth, nearly zero latency, and seamless integration, 7G will enable key applications such as data analysis, imaging, and artificial general intelligence. Speculations suggest that quantum computing could underpin 7G, but the probability of this happening is considered low. While quantum computing may be more powerful, it remains slower, and network evolution primarily relies on speed, making quantum computing less relevant in achieving the desired advancements in cellular technology.

In the quest for ever-evolving mobile network technologies, the emergence of 7G raises intriguing possibilities. With its envisioned capabilities, 7G could potentially revolutionize various industries, including healthcare, transportation, and entertainment. Seamless connectivity, real-time data processing, and advanced communication systems are expected to reshape the way we live, work, and interact with technology. As the demand for faster, more reliable, and intelligent networks continues to grow, the development of 7G holds the potential to unlock new opportunities and propel innovation to unprecedented levels, bringing us closer to a truly interconnected and digitally empowered society.

\section{Future references}
The evolution from 1G to 7G wireless networks has been a remarkable journey, marked by significant advancements in capacity, latency reduction, and technological innovation. Each generation of mobile networks has pushed the boundaries of communication, transforming the way we connect, communicate, and access information. From the introduction of 1G, which revolutionized mobile voice communication, to the lightning-fast speeds and seamless connectivity promised by 7G, the progression has been driven by a constant quest for improved performance and enhanced user experiences. 

The emergence of 7G as the next frontier in mobile network technology offers tremendous potential to address the ever-increasing demands of the digital era. With its dramatic increase in capacity and near-zero latency, 7G brings us closer to achieving our objectives of seamless, real-time communication and data processing. It promises to empower industries with transformative applications such as data analysis, imaging, and artificial general intelligence, revolutionizing sectors ranging from healthcare to transportation and entertainment. However, the realization of the full potential of 7G comes with significant challenges. The high costs associated with deploying such advanced technology pose financial considerations that need to be carefully addressed. While 7G represents a major milestone, it is important to recognize that it is not the endpoint of the evolutionary journey. The pursuit of cost efficiency will be a key driver for the next stage of network evolution, spurring further advancements in standards, technologies, and business models.

As we reflect on the trajectory of mobile network evolution, it is clear that each generation has built upon the achievements of its predecessor, propelling us towards increasingly sophisticated and connected digital landscapes. The transition from 1G to 7G has witnessed transformative shifts in capacity, latency, and network architecture, enabling unprecedented levels of communication and data exchange. The journey has been fueled by collaboration among industry stakeholders, visionary research, and a commitment to meet the evolving needs of a rapidly changing world.

In summary, the transition from 1G to 7G represents a remarkable continuum of advancements in mobile network technology. While 7G holds the promise of dramatically increased capacity, reduced latency, and revolutionary applications, we are reminded that the path to fully realizing its potential is still unfolding. Balancing technological advancements with cost efficiency will be crucial as we navigate the challenges and opportunities that lie ahead. The future of mobile networks, driven by the pursuit of seamless connectivity, real-time processing, and affordability, holds immense potential to transform industries and empower individuals in an interconnected and digitally empowered society.

\section{Conclusion}
The wireless mobile communication industry is experiencing rapid growth and continuous development. Over the years, cellular networks have evolved at an astonishing pace. This paper has aimed to comprehensively evaluate and trace the evolution of mobile networks, starting from the first generation (1G) and progressing to the fifth generation (5G). Additionally, we have discussed the future generations, including the anticipated advancements of 6G and the potential prospects of 7G.

Throughout this study, we have delved into the transformative changes brought about by each generation, from the introduction of voice communication in 1G to the revolutionary advancements in data rates, connectivity, and latency reduction in 5G. Moreover, we have explored the potential capabilities and applications expected in the upcoming generations, such as the increased bandwidth and the integration of intelligent technologies in 6G, as well as the remarkable improvements in capacity, latency, and transformative applications in 7G.

While this research provides valuable insights into the evolutionary path of cellular networks, it is important to acknowledge that the journey does not end with 7G. The wireless communication landscape will continue to evolve, fueled by advancements in technology, industry collaboration, and the ever-increasing demands of a digitally interconnected society.

Looking ahead, further research and exploration are necessary to fully understand and harness the potential of future generations of cellular networks. We encourage researchers and industry professionals to delve deeper into the subject matter, explore emerging technologies, and engage in interdisciplinary collaborations to unlock new possibilities and drive innovation in the field.

In conclusion, this paper has provided a comprehensive overview of the evolution of mobile networks, ranging from 1G to the forthcoming 7G. It highlights the transformative changes, technological advancements, and future prospects of wireless communication. As we continue to push the boundaries of connectivity and performance, the cellular network landscape will remain a dynamic and exciting domain, propelling us towards a more connected and digitally empowered future.

\end{document}